# Topological transitions in continuously-deformed photonic crystals

Xuan Zhu[1,¶], Hai-Xiao Wang[1,¶], Changqing Xu[1], Yun Lai[1], Jian-Hua Jiang[1,†], and Sajeev John[1,2,§]

[1]*College of Physics, Optoelectronics and Energy & Collaborative Innovation Center of Suzhou Nano Science and Technology, Soochow University, Suzhou 215006, China*

[2]*Department of Physics, University of Toronto, 60 St. George Street, Toronto, Ontario M5S 1A7, Canada*

[†]Email: jianhuajiang@suda.edu.cn, joejhjiang@hotmail.com

[§]Email: john@physics.utoronto.ca

[¶] These authors contributed equally.

**Abstract**:

We demonstrate that multiple topological transitions can occur, with high-sensitivity, by continuous change of the geometry of a simple 2D dielectric-frame photonic crystal consisting of circular air-holes. By changing the radii of the holes and/or the distance between them, multiple transitions between normal and topological photonic band gaps (PBGs) can appear. The time-reversal symmetric topological PBGs resemble the quantum spin-Hall insulator of electrons and have two counter-propagating edge states. We search for optimal topological transitions, i.e., sharp transitions sensitive to the geometry, and optimal topological PBGs, i.e., large PBGs with clean spectrum of edge states. Such optimizations reveal that dielectric-frame photonic crystals are promising for optical sensors and unidirectional waveguides.

I. **Introduction**

Photonic crystals (PCs) are the optical analogues of electronic semiconductors. In a PC, shape-tunable electromagnetic materials take the role of atoms or molecules in conventional crystals [1-3]. The periodic modulation of electromagnetic waves in real-space induces photonic bands and photonic band gaps (PBGs) with broad range applications in basic sciences and cutting-edge technology such as on-chip optical information processing, miniature optical devices, and subwavelength cavity quantum electrodynamics, to give but only a few examples [3]. In the past decade, certain prototype PCs have revealed a topological form of light-trapping: if two PBG materials of distinct band topology are put together, localized photonic modes will emerge and be guided on their mutual boundary, regardless of the specific geometry of the boundary [4-25]. Such light-confinement and wave-guiding are useful as robust and high-quality optical modes of almost arbitrary shape and extent. Since magneto-optical effects are typically weak and metals lead to significant loss in optical frequencies, all-dielectric photonic structures are highly preferred for optical-frequency applications [8,9], which are one of the main pursuits in future studies.

In this work, we propose a type of dielectric-frame two-dimensional (2D) PCs for transverse-electric (TE) photonic bands with time-reversal invariance. The PCs are comprised of triangular lattices of air-holes in a dielectric background, compatible with the current fabrication technology. With six air-holes in each hexagonal unit cell, our photonic structure is a generalization of the PC structure proposed in Ref. [19]. While the topological transition is revealed in Ref. [19] for transverse-magnetic (TM) modes by tuning the distance between the air-holes, here we perform a systematic study of topological transitions for TE modes with much richer geometry configurations. Our target here is to find optimal photonic structures for topological transitions and topological PBGs. A topological transition sensitive to the structure deformation can be used as optical sensors for biological and other applications. This function can be manifested in the transmission through the PC when the PBG is closed and reopen. On the other hand, an optimal topological PBG has a large gap-to-mid-gap ratio with a clean spectrum of edge states. In particular, for future synergy between topological photonic materials and optoelectronics/quantum optics, dielectric-frame PCs provide a feasible scheme.

In our calculations, we use GaP ($\varepsilon = 10.2$), silicon ($\varepsilon = 12$), or germanium ($\varepsilon =$

16) as the dielectric background. A topological PBG as large as 20%, with a clean spectrum of helical edge states, is demonstrated. The switching between the topological and normal PBGs takes place as the radii and positions of the air holes are continuously tuned. Contrary to the deformed-honeycomb [19,23] and core-shell [20] PCs, our dielectric-frame PCs exhibit *multiple* topological transitions when the radii and/or positions of the air-holes are varied continuously. The rich phase diagram enables effective tuning of the photonic properties, which may have potential applications as highly-sensitive optical sensors.

This paper is organized as follows. In Sec. II we present the design and architecture of the dielectric-frame PCs and the properties of their photonic bands. In section III, we study multiple topological transitions and their connection to the PC geometry. In section IV, we describe the nature of the topological PBGs and their protected edge states. In Sec. V we discuss possible optical sensors based on the geometry-sensitive topological transitions. We conclude in Sec. VI.

## II. Dielectric-Frame Photonic Crystals

We study 2D PCs of triangular types of lattices illustrated in Fig. 1. The lattice vectors are $\vec{a}_1 = (a, 0)$ and $\vec{a}_2 = (\frac{a}{2}, \frac{\sqrt{3}a}{2})$ with the lattice constant $a$. In each hexagonal unit cell, there are six circular air-holes in a dielectric background. The six air-holes are of identical radius $r$ [see Fig. 1(a)]. The lattice and configurations of air-holes have the $C_6$ rotational symmetry. The tunable geometric parameters are the radius $r$ of the air-holes and the distance between the adjacent air-holes within a unit-cell $l$. The latter is also equal to the distance between an air-hole and the unit-cell center. By tuning the radius $r$ and the distance $l$, the $C_6$ symmetry is preserved, while various geometries can be realized as illustrated in Fig. 1. There are different scenarios for the geometry transitions. For small air-holes [see Fig. 1(a)], the main geometry transition takes place at $l = a/3$ where a honeycomb lattice of air-holes is realized. As found in Ref. [19], at $l = a/3$ the sub-lattice symmetry in the honeycomb lattice enforces the emergence of the double Dirac point. For $l < a/3$ or $l > a/3$, the double Dirac point is gapped and a PBG emerges. It was found that the PBG for $l < a/3$ is of trivial topology, while the PBG for $l > a/3$ is of nontrivial topology which resembles the quantum spin-Hall insulator of electrons [19]. We refer to the lattices with $l < a/3$ as type-I triangle lattices, while the lattices with $l > a/3$ are called type-II triangle lattices. Another interesting limit is $l = a/2$ [see the fourth panel of Fig. 1(a)], where the structure becomes a Kagome lattice. These are lattices of air-holes with small radii for various distance $l$.

If the radii of the air-holes are large, different geometries and transitions may appear. As illustrated in Fig. 1(b), starting from small $l$, where the lattice is a triangle lattice of a small cluster of air-holes, the first geometric transition takes place at $l + r = a/2$. For $l + r \geq a/2$, air-holes from adjacent unit-cells start to overlap and the lattice becomes a honeycomb-triangle lattice of dielectric materials. That is, there is a honeycomb lattice of dielectric materials at the corners of the unit cell, together with a triangular lattice of dielectric material at the unit-cell center [see the second panel of Fig. 1(b)]. As the distance $l$ increases, such a geometry transition can take place before the transition into the honeycomb lattice of air-holes, if the radius is large, $r > a/6$. Besides, for such large air-holes, the honeycomb lattice of air-holes is equivalent to a triangle lattice of dielectric materials, as shown in the third panel of Fig. 1(b). At the critical value of $r = a/6$, the air-holes in the honeycomb lattice start to overlap with each other. Therefore, the geometric transitions are different for large radii $r > a/6$, compared to small radii $r < a/6$. As revealed in this study, the topological transitions are also fundamentally different between PCs with small and large air-holes.

Before going into details, we provide two concrete examples of dielectric-frame PCs with PBGs of trivial and nontrivial topology. Photonic band diagrams of two PCs for the TE harmonic modes are presented in Fig. 2, where we choose $r = 0.2a$ for two different distances: $l = 0.275a$ for Fig. 2(a), and $l = 0.475a$ for Fig. 2(b). We use the commercial software COMSOL Multi-physics to calculate the photonic band structures. Throughout this paper, we mostly use the dimensionless angular frequency $\frac{\omega a}{2\pi c}$, where $c$ is the speed of light in vacuum. In addition, the default length unit is the lattice constant $a$. The Brillouin zone of the PC is illustrated in the inset of Fig. 2(a). The photonic bands at the Γ point is doubly degenerate for both the dipole and quadrupole modes. The magnetic field profiles in Fig. 2(b) indicate that the dipole modes are below the quadrupole modes for $l = 0.275a$. We shall term such a band-order as the normal band-order. Fig. 2(d) shows that, for $l = 0.475a$, the dipole modes are *above* the quadrupole modes at the Γ point. Such a band-order is called as the parity-inversed band-order. As found in Refs. 19 and 20, the parity-inversed band-order corresponds to nontrivial band topology. We elaborate on the nature of this band topology in Sec. IV.

The topological transition takes place when the *p* and *d* bands become degenerate, where a double Dirac point emerges at the Γ point. Band-gap closing and reopening through Dirac points is a signature for a topological transition. For 2D PC, this scenario holds whenever the PC has $C_6$ rotation symmetry [18-20]. The underlying mechanism is that the $C_6$ point group has two doubly degenerate representations, which are isomorphic to the *p* and *d* modes [see Fig. 2] [18-20]. These are manifested as the $p_x$, $p_y$, $d_{x^2-y^2}$ and $d_{xy}$ modes which can form four pseudo-spin states,

$$|p_+\rangle = |p_x + ip_y\rangle, \qquad |p_-\rangle = |p_x - ip_y\rangle,$$
$$|d_+\rangle = |d_{x^2-y^2} + id_{xy}\rangle, \qquad |d_-\rangle = |d_{x^2-y^2} - id_{xy}\rangle. \qquad (1)$$

Moreover, the *p* doublet and the *d* doublet have opposite parities, which ensures the *k*-linear couplings between them, resembling the Dirac equation, as shown in Refs. [19,20]. Such an analog with the Dirac equation for electrons connects the parity-inversed photonic band-order with the quantum spin Hall effect of electrons [19].

**III. Multiple Topological Transitions**

In Fig. 2 the order of the *p* and *d* bands are modified by tuning the distance *l*, as was found in Ref. [19] for the TM modes. We now show that the phase diagram is much richer than the phase diagram revealed in Refs. [19,20]. In Fig. 3 the phase diagram is shown for both small and large radii of the air-holes with varying distance *l*. The band-order between the *p* and *d* bands is characterized by the following quantity [20],

$$\omega_{pd} = \frac{2(\omega_d - \omega_p)}{\omega_d + \omega_p}, \qquad (2)$$

where $\omega_d$ and $\omega_p$ are the frequencies of the *d* and *p* bands at the Γ point, respectively. For small radius $r = 0.13a$, there is indeed a transition between normal and parity-inversed band-orders at $l = a/3$, which is the topological transition found in Ref. [19]. In addition to this transition, there is another transition at smaller distance $l = 0.25a$. Thus, for smaller *l*, the *p* bands become higher in frequency than the *d* bands, leading to parity-inversed band-order. From the field-profiles in Fig. 3(b), one can see that the switch of band-order is induced by the transition from air-like modes to dielectric-like modes triggered by the geometric deformations. Specifically, for a type-II triangle lattice, the *p* modes are focused in the dielectric region (dielectric-like modes), while the *d* modes concentrate in the air region (air-like modes). As the distance $l$ reduces, the PC transform into the type-I triangle lattice, while the *d* modes become dielectric-like and the *p* modes become air-like. For smaller distance $l < 0.25a$, the air-holes go to the center of the unit-cell. In this regime, both *d* and *p* modes become dielectric-like and are focused on the outer region of the unit-cell. During these processes, the frequency of the *d* and *p* modes are strongly modified, since the frequency and the field-profiles of these eigen-modes are closely related to each other.

Very different from the small radius regime, there are multiple topological transitions with sizeable PBGs when the radius of the air-holes is large. In Fig. 3(c) we show the phase diagram for $r = 0.2a$. There are *four* topological transition points. One of them is the honeycomb lattice limit $l = a/3$. These transitions are triggered by the abundant geometric transitions. From the field-profiles, one can see that the *p* modes go from dielectric-like (in the outer region) to air-like and then to dielectric-like (in the inner region), as the distance $l$ increases from 0 to $a/2$. The *d* modes, which always concentrate in the outer region, transform from dielectric-like to air-like. These transitions strongly modulate the band-order, as revealed by the dependence of $\omega_{pd}$ on the distance $l$. We remark that the significant modifications of the field-profiles and band-order induced by geometric transitions are unique features in our dielectric-frame PCs. Such phenomena were *not* found in previous studies on core-shell [20] or deformed-honeycomb [19] PCs for TM modes. The transitions between air-like and dielectric-like modes induce significant changes in the frequencies of the *p* and *d* bands, leading to very large PBG with nontrivial topology.

A complete PBG between the *p* and *d* bands is necessary to realize the photonic analog of quantum spin Hall insulator, in addition to the parity-inverted band-order. From Figs. 3(a) and 3(c), the largest PBG with nontrivial topology can be realized at $l = 0.4a$ and $r = 0.13a$ with gap-ratio close to 20%. Here the gap-ratio $|\Delta\omega|$ is defined as the ratio of the gap size over the mid-gap frequency. To elaborate on these topological transitions, we plot the photonic band-structures at the four transition points except that with $l = a/3$ (the honeycomb limit). Figs. 4(a), 4(b), and 4(c) show that in all these transition points double Dirac points emerge at the $\Gamma$ point as consequence of the degeneracy between the *p* and *d* bands. A clean spectrum of a double Dirac point exists for a large frequency range for the parameters $l = 0.4a$ and $r = 0.2a$. Fig. 4(d) shows the zoom-in 3D dispersion around the double Dirac point, which exhibits perfect conic dispersion.

To consolidate our picture of topological transitions, we further study the phase diagram by varying the radius $r$ for fixed distance $l$. Fig. 5 shows that topological transitions can also be triggered by tuning the radius $r$ for both $l = 0.4a$ and $l = 0.5a$. These findings are *in contrary to common beliefs* that the topological transition is triggered by tuning the distance $l$ through the Dirac points in the honeycomb lattice [19,23,24,25]. For both $l = 0.4a$ and $l = 0.5a$ (Kagome lattice), the topological PBGs are realized for small radii. For $l = 0.4a$, the transition between the topological PBG and normal PBG can be realized by continuous tuning of the radius around $r = 0.2a$. The drastic change of the PBG around this point, which was not reported before, can be useful for highly tunable or sensitive optical devices. For example, it might be used for designing optical sensors, since the attachment of biomarkers to the dielectric surfaces is nearly equivalent to tuning the radius of the air-holes [26,27]. We also notice that the largest PBG with nontrivial topology is realized at $r = 0.16a$ where the gap-ratio is nearly 20%. This large PBG is ideal for strong topological light-confinement and protected light-propagation along the boundaries between normal and topological PBG materials.

**IV. Nature of Topological Photonic Band Gaps and Their Protected Edge States**

To elucidate the electromagnetic properties of the topological structures and transitions, we now derive the effective "Hamiltonian" for the photonic spectrum near the double Dirac points using the $k \cdot p$ perturbation theory. The Bloch functions must satisfy the following eigenvalue equation from the Maxwell equations,

$$\nabla \times \frac{c^2}{\varepsilon(\vec{r})} \nabla \times \vec{H}_{n\vec{k}}(\vec{r}) = \omega_{n\vec{k}}^2 \vec{H}_{n\vec{k}}(\vec{r}). \qquad (3)$$

Here $c$ is the speed of light in vacuum, $\varepsilon(\vec{r})$ is the position-dependent relative permittivity, $\vec{H}$ is the magnetic field, while the indices $n$ and $\vec{k}$ label the band number and the wavevector, respectively. Following the procedures in Refs. [20,21], we expand the Bloch functions at finite (but small) $k$ with the Bloch function at the BZ center. By restricting our analysis only on the four *p* and *d* states, we obtain a 4× 4 matrix representation of the photonic Hamiltonian. Although this analysis is valid only when the four *p* and *d* states are close to each other and far-away from other bands, it is adequate to reveal the nature of topological transitions and the band topology in our PCs. We work with the basis of $(|d_+>, |p_+>, |d_->, |p_->)^T$. After some tedious calculations [20,21], we obtain the following matrix form of the photonic eigenvalue equation,

$$\begin{pmatrix} \omega_d^2 & \alpha k_+ & 0 & 0 \\ \alpha^* k_- & \omega_p^2 & 0 & 0 \\ 0 & 0 & \omega_d^2 & \alpha k_- \\ 0 & 0 & \alpha^* k_+ & \omega_p^2 \end{pmatrix} \begin{pmatrix} c_{d_+} \\ c_{p_+} \\ c_{d_-} \\ c_{p_-} \end{pmatrix} = \omega_{n\vec{k}}^2 \begin{pmatrix} c_{d_+} \\ c_{p_+} \\ c_{d_-} \\ c_{p_-} \end{pmatrix}, \qquad (4)$$

which is valid up to the linear-order in $\vec{k}$. Here $\omega_d$ and $\omega_p$ are the frequency of the $d$ and $p$ bands at the Γ point, respectively, $k_\pm = k_x \pm ik_y$, and $\alpha$ is the $k \cdot p$ coefficient. Time-reversal symmetry requires that $\alpha$ is a purely imaginary number. Here $c_j$ ($j = d_+, p_+, d_-, p_-$) are the coefficients for the eigenstates. We note that the above equation resembles the Hamiltonian eigenvalue equation that describes the quantum spin Hall effect of electrons in Refs. [28,29,30]. The analog with the quantum spin Hall effect of electrons indicates that the parity-inversed band-order $\omega_d < \omega_p$ is equivalent to the quantum spin Hall insulator, while the normal band-order $\omega_d > \omega_p$ has trivial band topology. In fact, the construction of double degeneracy and the parity-inversion are the two crucial prerequisites for the simulation of Dirac physics and band topology in photonics [21,22]. In 3D PCs, a $C_6$ rotation symmetry together with inversion symmetry in hexagonal lattices [21], or two orthogonal screw symmetries in tetragonal lattices [22], can create 3D photonic Dirac points.

A fundamental evidence of the band topology is the appearance of protected edge states at the interfaces between two topologically distinct PBG structures. In Fig. 6 we show that there are helical edge states in the common PBG when two PCs of different band topology are placed together. The band structures of the bulk and edge states are calculated via a supercell geometry illustrated on the top of Fig. 6(a). On each boundary, there are two edge states of opposite velocity [see Figs. 6(a) and 6(b)]. Their spectra are related by time-reversal symmetry,

$$\omega_A(k) = \omega_B(-k).$$

To illustrate the nature of the edge states, we plot the profiles of the (time-averaged) Poynting vectors $\vec{S} = \frac{1}{2} Re[\vec{E} \times \vec{H}^*]$ and the magnetic field $H_z$ (the imaginary part) together in Fig. 6(c) for the two edge states with the same frequency but opposite wavevectors, i.e., a Kramers pair. It is clear from the opposite vorticity of the Poynting vectors that these two edge states have opposite orbital angular momenta (OAM). Specifically, the OAM is along the *z* direction and can be defined as $L_z = xP_y - yP_x$ where the momentum of the electromagnetic wave is related to the Poynting vector as $\vec{P} = \vec{S}/c$. Here the origin of the coordinate is taken as the center-of-mass of the edge states in a supercell calculation, namely that $\int d\vec{r} w(\vec{r}) \vec{r} = 0$ with $\vec{r}$ being the 2D coordinate vector and $w(\vec{r}) = \frac{1}{2}(\varepsilon_0 \varepsilon(\vec{r}) |\vec{E}|^2 + \mu_0 \mu(\vec{r}) |\vec{H}|^2)$ is the energy density. We shall term the edge states with positive OAM along the *z* direction as the "spin-up" state, while the negative OAM states are denoted as the "spin-down" state. The spin-up and spin-down edge-photons have opposite wavevectors (group velocities), which is termed as the "spin-wavevector locking" in the language of the electronic quantum spin Hall effect. The emergence of finite OAM for a photonic Kramers pair and the "spin-wavevector locking" [29,30] in purely dielectric materials in the linear-response regime is quite striking. This unique property can be utilized to excite OAM beams with a point-like source on the edge: the positive OAM state goes to the right-hand side, while the negative OAM state goes to the left-hand side.

    We now demonstrate the topological protection of the edge states through unidirectional propagation of the spin-up (spin-down) edge states in a *Z*-shaped boundary. The spin-up (spin-down) edge states can be selectively excited via a specially designed point-like source carrying OAM +1 (-1), as illustrated in Ref. [24]. The simulation set-up for unidirectional propagation of edge states is shown in Fig. 7(a). A *Z*-shaped boundary between two PCs of different band topology is placed in a box geometry surrounded by perfectly matching layers (PMLs) in all directions. The electromagnetic waves excited by the point-like source (denoted as the yellow star) propagate along the *Z*-shaped boundary, which eventually leave the PC region and are absorbed by the PMLs. The PC with trivial band topology is chosen as $l = 0.4a$, $r = 0.21a$, and dielectric constant $\varepsilon = 16.8$. The PC with nontrivial band topology is set as $l = 0.4a$, $r = 0.16a$ and $\varepsilon = 10.2$. This set-up gives a common PBG between the two PCs in the range of $(0.68 \frac{2\pi c}{a}, 0.7 \frac{2\pi c}{a})$. We use COMSOL Multiphysics to perform the full wave simulation. From Figs. 7(b) and 7(c), we can see that the spin-up edge states propagate along one direction, while the spin-down edge states

along the opposite direction. There is no visible back-reflection when the edge modes go through the corner-turning in the Z-shaped boundary, demonstrating robust unidirectional transport and topological protection.

**V. Optical Sensors Based on Topological Transitions**

In this section we demonstrate that the sharp topological transition can be used for optical sensors. In particular, we use the topological transition at $l = 0.4a$ and $r = 0.2a$ to demonstrate the sensitivity of the photonic transmission spectrum on the thickness of the biomarkers attached to the dielectric interfaces. Using advanced schemes of bio-recognition such as antibody-antigen binding and DNA aptamer-protein binding, targeted biomarkers can be attracted to the dielectric-air interfaces in the testing processes [26,27]. The amount of biomarkers are quantified by their thickness $\delta r$, as schematically shown in the inset of Fig. 8(a). The radius of the air-holes are then reduced to $r - \delta r$. The refractive index of the biomarkers is taken as 1.5 as in Refs. [26,27].

We first present the dependences of frequencies of the $p$ and $d$ bands at the $\Gamma$ point, $\omega_p$ and $\omega_d$, on the thickness of the biomarker layers, $\delta r$. The yellow and cyan regions represent normal and topological PBGs, respectively. There are two topological transition points in Fig. 8(a), which correspond to the emergences of the Dirac points while the PBG is closed and then reopen. Such topological transitions are found to be very sensitive to the thickness of the biomarker layers. As shown in Fig. 8(a), both frequencies, $\omega_p$ and $\omega_d$, change significantly with the thickness of the biomarker layers.

Fig. 8(b) shows the schematic picture for the structure of the optical sensor. Transmission spectrum through the PC region reflects the change of the photonic band structure induced by the biomarkers. In Fig. 8(c) the transmission spectra for three different thicknesses of the biomarker layer, $\delta r/a = 0, 0.05, 0.1$, are presented. Distinct transmission spectrum are shown for the frequency window $(0.64, 0.7)\frac{\omega a}{2\pi c}$. The first case, $\delta r = 0$, shows a transmission spectrum where the low frequency window, $\frac{\omega a}{2\pi c} < 0.645$, touches the lower band edge of the $d$ bands. For the second case, $\delta r = 0.05a$, the sinusoidal transmission spectrum represents photon propagation through the bulk bands. The third case, $\delta r = 0.1a$, has a transmission gap, $(0.654, 0.672)\frac{\omega a}{2\pi c}$, representing the complete PBG due to reversed band-order. The sensitive change of the transmission spectrum due to the small variation of the thickness of biomarker layer indicates potential applications of our PC as optical sensors.

## VI. Conclusions and Outlook

In conclusion, we have demonstrated that carefully designed 2D dielectric-frame PCs can exhibit large PBGs with nontrivial band topology. The resulting electromagnetic behavior is analogous to the quantum spin Hall effect of electrons, in which photonic OAM plays the role of electronic spin. These photonic analogues of quantum spin Hall insulators, consisting of non-absorbing dielectric materials, may be suitable for tunable, high-sensitivity, optical-frequency, on-chip devices. We find that *multiple* transitions between trivial and topological photonic bands can be achieved by merely tuning the geometry of the air-holes. The underlying mechanism is the geometric switching between different mode profiles (air-like and dielectric-like modes). Complete topological PBGs as large as 20% can be achieved. Helical edge states are found in the common PBG of two PCs with different band topology, where a Kramers pair of edge states are shown to have opposite OAMs and wave-vectors. Robust unidirectional edge-state propagation is demonstrated via full-wave simulation. Our findings illustrate that dielectric-frame PCs can provide highly-tunable and sensitive photonic chips with topological phenomena at optical-frequency in the subwavelength regime. We also show that the geometry-sensitive topological transitions can be exploited for optical sensors.

**Acknowledgements**

XZ, HXW and JHJ acknowledge supports from the National Natural Science Foundation of China (NSFC Grant No. 11675116) and the Soochow University. CX and YL thank supports from National Natural Science Foundation of China (Grant Nos. 61671314 and 11374224). SJ acknowledges support from Natural Sciences and Engineering Research Council of Canada.

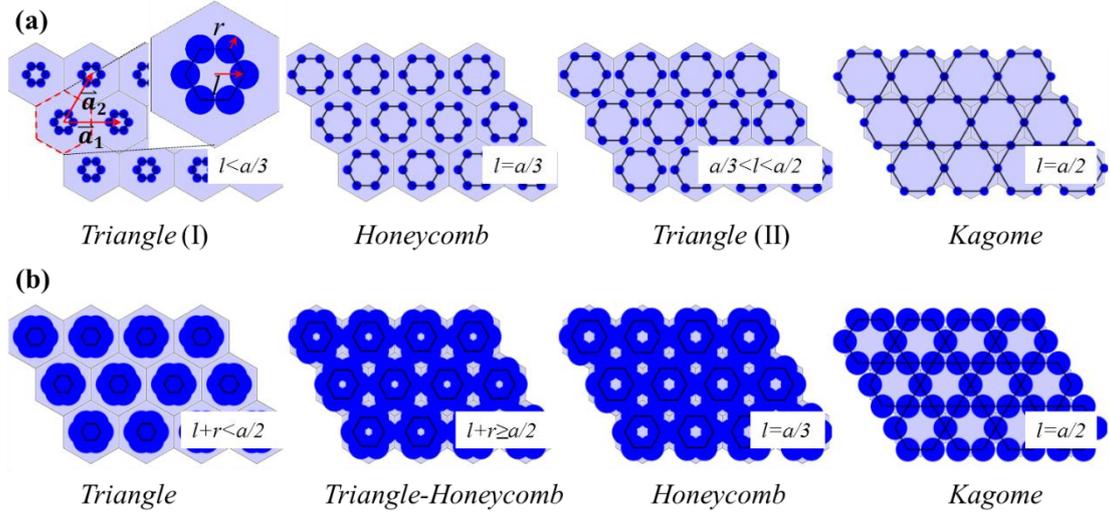

Fig. 1. Geometric transitions in the dielectric-frame 2D PCs. Each hexagonal unit-cell has six air-holes (dark-blue regions) in the dielectric background (gray-blue regions). Each air-hole has radius $r$. The distance between the center of an air-hole and the center of the unit-cell is $l$. The structure has $C_6$ symmetry and the lattice constant is $a$. By tuning the radius $r$ and distance $l$, the PC can experience various geometric transitions. (a) For air-holes of small radius, the main transition is between two types of triangle lattices: one has $l < a/3$ ("triangle I"), while the other has $l > a/3$ ("triangle II"). The transition takes place at $l = a/3$ where the PC becomes a honeycomb lattice of air-holes. A special limit is $l = a/2$, where the PC is manifested as a Kagome lattice. (b) For air-holes of large radius, the transition is different: When $l + r < a/2$, the PC is a triangle lattice of air-holes. For larger distance $l + r \geq a/2$, the PC becomes a triangle-honeycomb lattice of dielectrics. In the special cases of $l = a/3$ and $l = a/2$, the PCs are still the honeycomb and Kagome lattices of air-holes, respectively.

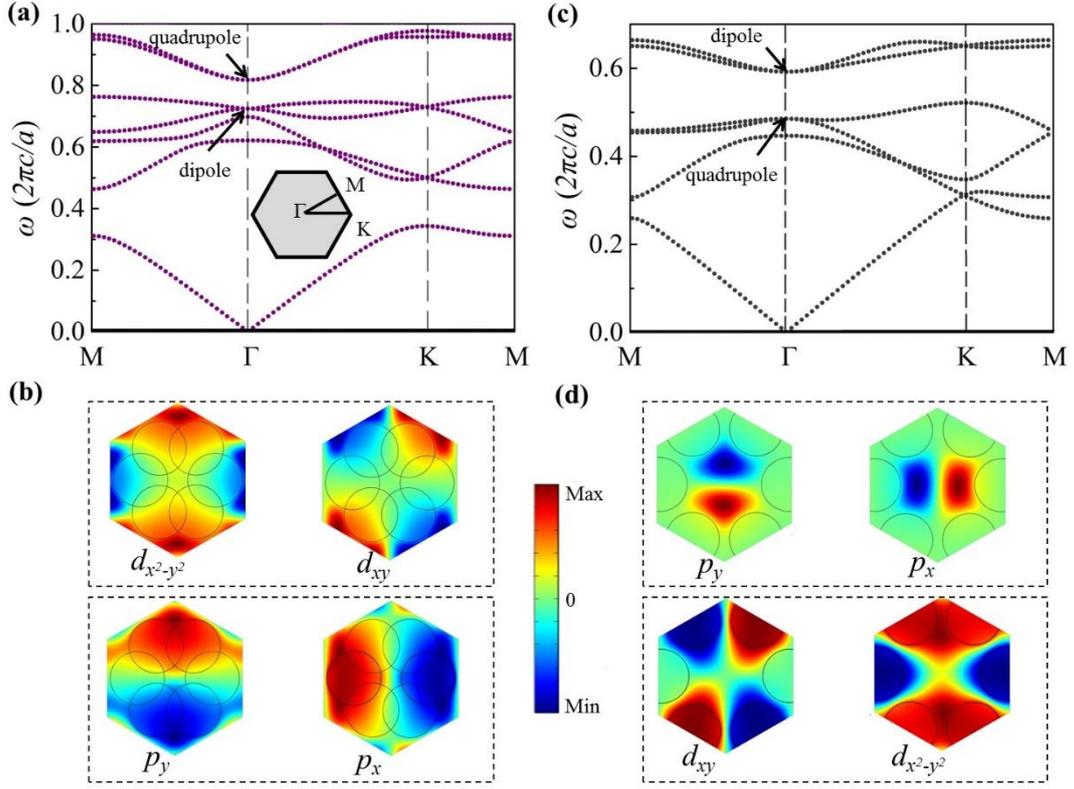

Fig. 2. Band diagrams and eigenmodes in dielectric-frame 2D PCs with normal and inversed band-order. The dielectric background is made of GaP. (a) Band diagram for the 2D PC with $r = 0.2a$ and $l = 0.275a$. The dipole modes have lower frequency than the quadrupole modes at the $\Gamma$ point. (b) The magnetic field $H_z$ profiles of dipole and quadrupole modes at the $\Gamma$ point. The six circles in each unit cell represent the air-holes. (c) Band diagram for the 2D PC with $r = 0.2a$ and $l = 0.475a$. The dipole modes have higher frequency than the quadrupole modes at the $\Gamma$ point. (d) Magnetic field $H_z$ profiles of dipole and quadrupole modes at the $\Gamma$ point. In (b) and (d), the upper panels represent modes of higher frequency, while the lower panels are for modes of lower frequency.

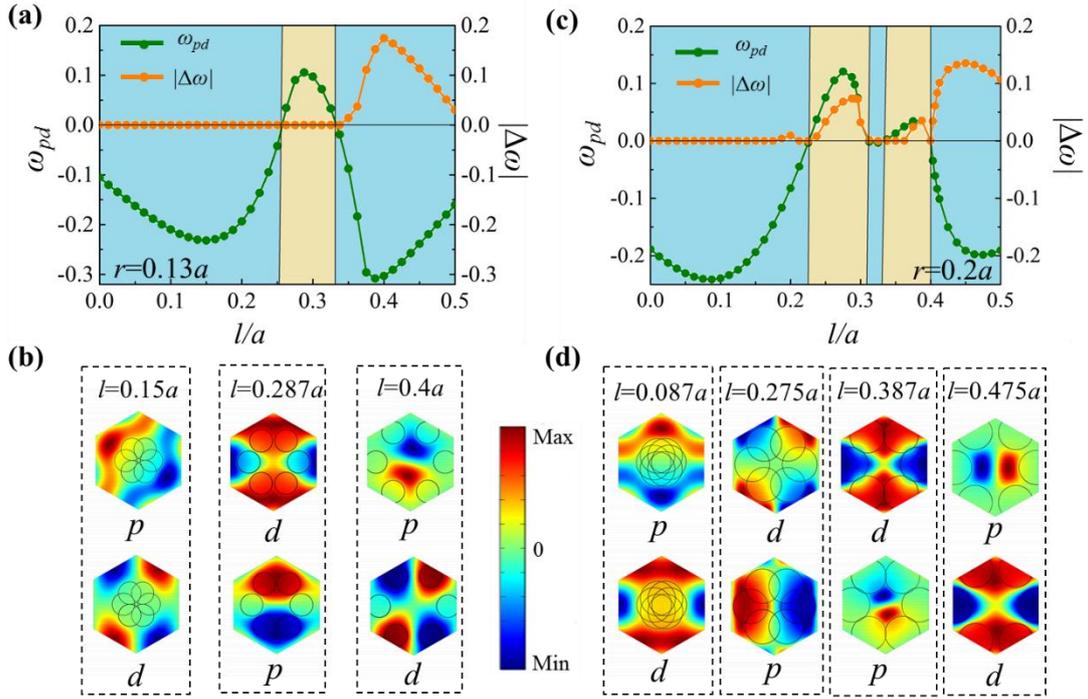

Fig. 3. (a) Phase diagram of parity-inversion induced topological transitions for a small radius $r = 0.13a$ with varying distance $l$. The phase transition is characterized by the quantity $\omega_{pd}$ (represented by the green curve). The cyan regions stand for the reversed band-order, while the yellow region denotes normal band-order. The orange curve represents the size of the complete PBG between the $p$ bands and the $d$ bands. (b) Magnetic field $H_z$ profiles of dipole and quadrupole modes at the BZ center for $r = 0.13a$ and various $l$. (c) Phase diagram of parity-inversion induced topological transitions for a large radius $r = 0.2a$ with varying distance $l$. (d) Magnetic field $H_z$ profiles of dipole and quadrupole modes for $r = 0.2a$ and various $l$. In Figs. (b) and (d) the upper (lower) field-profiles are for higher (lower) frequency modes. The field-profiles of the degenerate $d$ ($p$) bands are adopted from one mode out of the $d$ ($p$) doublets.

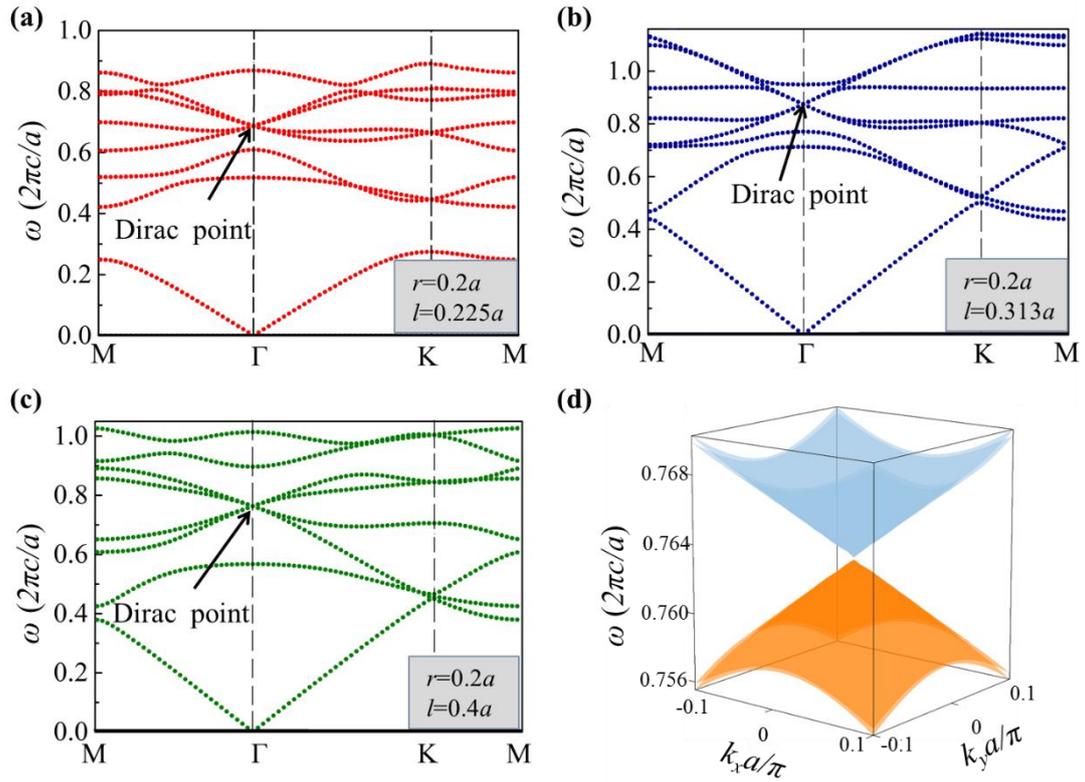

Fig. 4. Double Dirac points emerging at the topological phase transition points for $r = 0.2a$ and various distance $l$. (a)-(c) Band diagrams. (d) Zoom-in dispersion of the double Dirac point in (c).

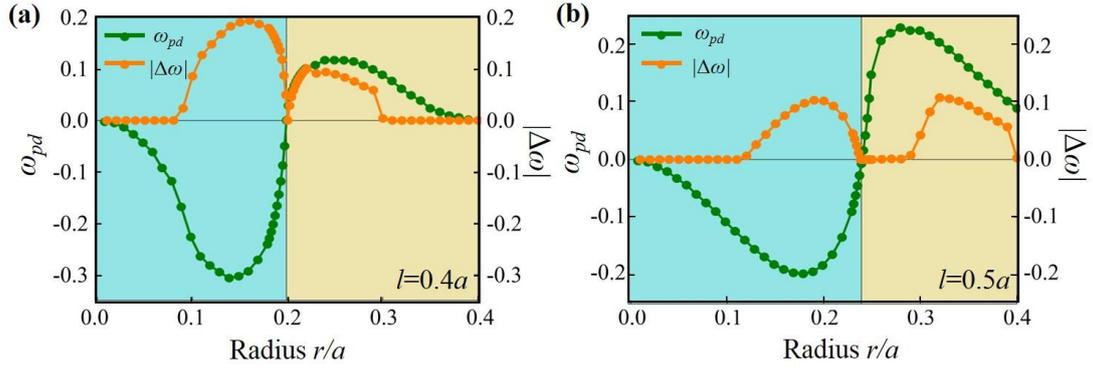

Fig. 5. Phase diagram of the dielectric-frame PCs for (a) $l = 0.4a$ and (b) $l = a/2$ for various radii $r$. (b) represents the case of a Kagome lattices of air-holes. The cyan (yellow) regions stand for the reversed (normal) band-order, i.e., $\omega_{pd} < 0$ ($\omega_{pd} > 0$). The orange curve represents the size of the complete PBG between the $p$ bands and the $d$ bands. Significant complete PBGs can be achieved in both (a) and (b).

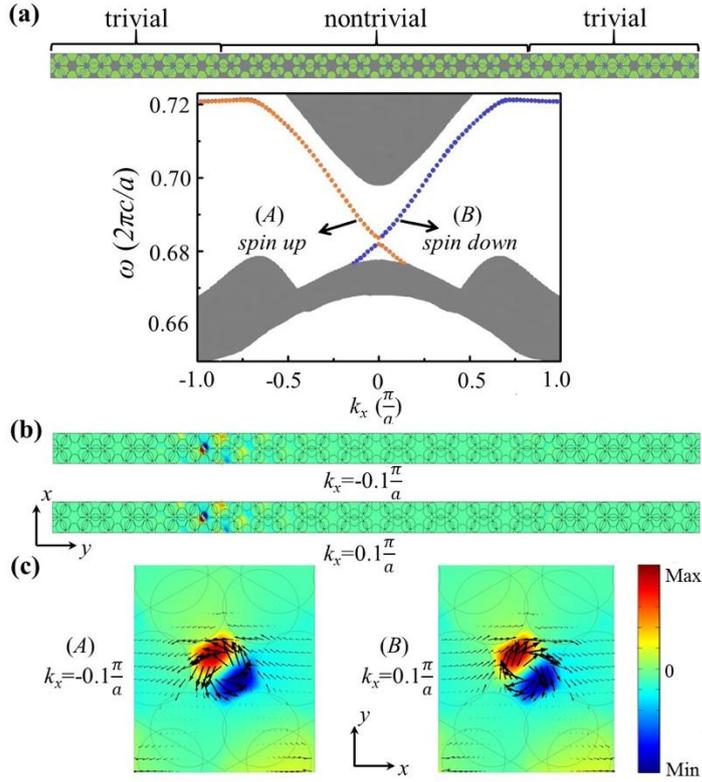

Fig. 6. Topological edge states of photons. (a) Band diagram for a supercell calculation of the edge states. The gray regions represent the bulk photonic bands, the orange and blue curves stand for the edge states. The upper panel shows the geometry of the supercell. (b) Magnetic field $H_z$ profiles of the two edge states at the same frequency $\omega = 0.688(\frac{2\pi c}{a})$ for opposite wavevectors. (c) Zoom-in graph of the magnetic field profiles for the same edge states as in (b). Poynting vector profiles are also included in (c), as represented by the black arrows.

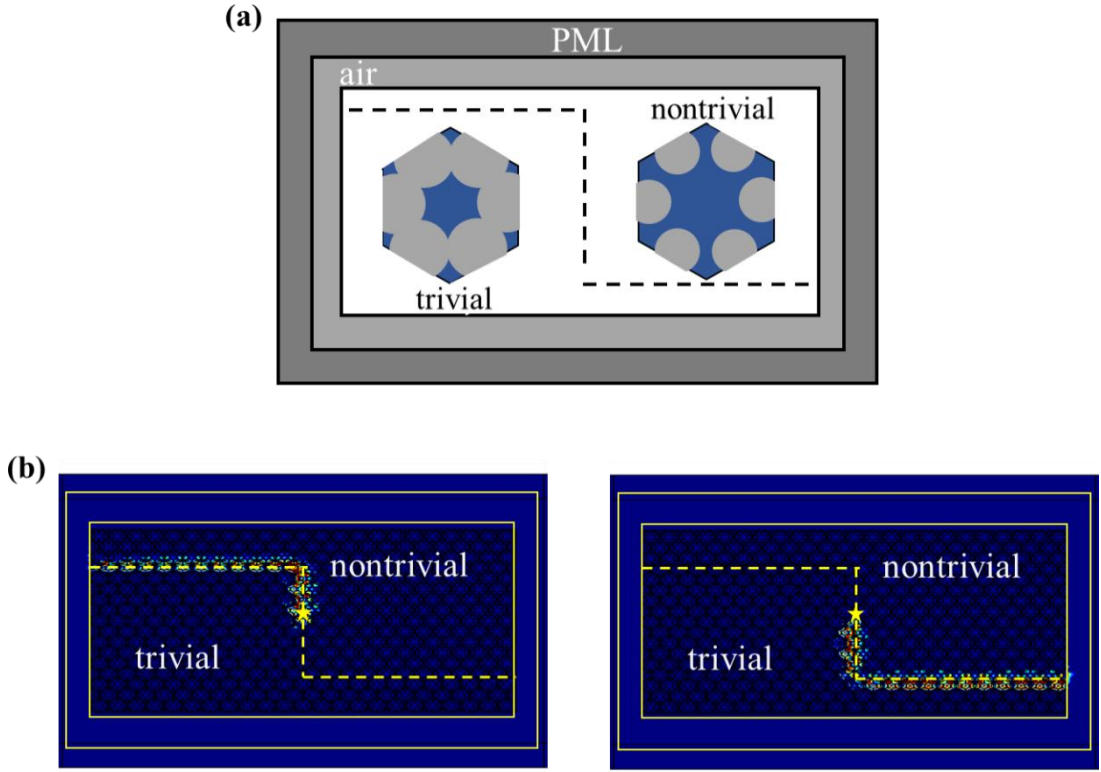

Fig. 7. Photonic edge states exhibit unidirectional propagation along *Z*-shaped boundaries between PCs with complete PBGs of trivial and nontrivial topology. (a) Illustration of the computational cell. The horizontal section of the *Z*-shaped boundary consists of zigzag boundaries between the two PCs. (b) Unidirectional propagation of photons excited by a point source of orbital angular momentum -1, corresponding to the spin-down edge states. (c) Unidirectional propagation of photons excited by a point-like source of orbital angular momentum +1, corresponding to the spin-up edge states. Electromagnetic waves excited by the source are eventually absorbed by the PMLs surrounding the whole structure.

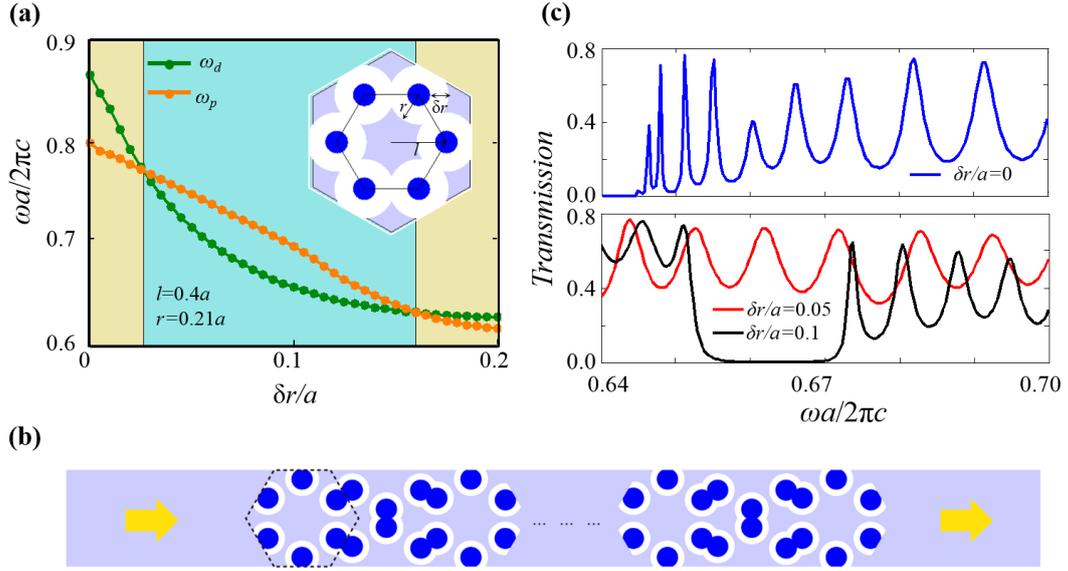

Fig. 8 (a) Frequencies of the $p$ and $d$ bands at the $\Gamma$ point, $\omega_p$ and $\omega_d$, as functions of the thickness of the attached biomarker layer, $\delta r$. The yellow and cyan regions represent normal and topological PBGs, respectively. The structure of the unit-cell is illustrated in the inset. The dark-blue regions are the air-holes. The white region represents the biomarker attached to the dielectric background (the gray-blue regions). The original radius of the air-hole is $r = 0.21a$. The distance between the center of an air-hole and the center of the unit-cell is $l = 0.4a$. (b) Illustration of photon transmission through a PC region with 20 unit-cells. The gray-blue regions represent the dielectric material. The white regions denote the biomarkers, while the dark-blue regions stand for air-holes. (c) Photon transmission spectrum for various thickness of the biomarkers. The blue, red and black curves correspond to $\delta r/a = 0, 0.05, 0.1$, respectively.